\documentclass[aps,pra,superscriptaddress,twocolumn,showpacs,amsmath,amsfonts,amssymb,floatfix]{revtex4}
\usepackage{enumerate}
\usepackage{mathtools}
\usepackage{amsmath}
\usepackage{graphicx}
\usepackage{epsfig}
\usepackage{sidecap}
\usepackage{hyperref}
\usepackage{color}
\usepackage{enumerate}
\usepackage{bm}

\begin{document}

\title{Quantum State Transfer through Coherent Atom-Molecule Conversion in Bose-Einstein Condensate}

\author{Subhrajit Modak}
\affiliation{Indian Institute of Science Eduaction and Research Kolkata, Mohanpur, West Bengal - 741246, India}

\author{Priyam Das} \thanks{Corresponding author: P.D. - daspriyam3@gmail.com}
   \affiliation{Department of Physics, Indian Institute of Technology Delhi, Hauz Khas, New Delhi - 110016, India }

\author{Prasanta K. Panigrahi}
   \affiliation{Indian Institute of Science Eduaction and Research Kolkata, Mohanpur, West Bengal - 741246, India}

\begin{abstract}
We demonstrate complete quantum state transfer of an atomic Bose-Einstein condensate to molecular condensate, mediated by solitonic excitations in a cigar shaped mean-field geometry. Starting with a localized solitonic atomic condensate, we show compatible gray solitonic configuration in the molecular condensate, which results in complete atom-molecule conversion. The effect of inter and intra-species interactions on the formation of molecular condensate is explicated in the presence of Raman Photoassociation. It is found that photoassociation plays a crucial role in the coherent atom-molecule conversion as well as in the soliton dynamics. The gray soliton dispersion reveals bistable behaviour, showing a re-entrant phase in a physically accessible parametric domain.
\end{abstract}


\maketitle

\section{Introduction}

Controlled formation of molecular Bose-Einstein condensate (BEC) is an area of active research \cite{hutson2006molecule}. The conversion of macroscopic number of quantum degenerate atoms into molecular dimers \cite{wynar2000molecules,mark2005efficient} and other poly-atomic structures \cite{refId0} have been observed in several experiments \cite{xu2003formation,herbig2003preparation,winkler2007coherent,danzl2008quantum}. Due to complex spectral pattern of rovibrational levels, molecules cannot be cooled directly using laser cooling techniques. Interestingly, the  richer energy level structure of the molecules offer more possibilities for coherent control of molecular dynamics. A promising route towards a molecular condensate is the conversion of atomic BEC into molecules \cite{xu2003formation,herbig2003preparation,mckenzie2002photoassociation} in ultracold condition. Photoassociation (PA) \cite{ulmanis2012ultracold,roy2016photo,naskar2017electromagnetically} and  Feshbach resonance (FR) \cite{tasgal2006molecule} are two widely employed methods for connecting ultracold atoms to molecular bound states. PA has the added advantage of  frequency tunability, which can be used for selecting a particular rovibrational ground state. The interacting assembly of such homonuclear or heteronuclear molecules, with bosonic or fermionic atoms in their condensed phase, has been termed as `superchemistry' \cite{PhysRevLett.84.5029,moses2017new}. Collective oscillations between the atomic and molecular condensates characterize this ultracold chemical dynamics, not governed by the classical Arrhenius law \cite{1367-2630-17-5-055005}.  Novel quantum effects, such as super-selectivity in dissociating triatomic molecules and the quantum size effect, have been identified in this regime \cite{1367-2630-17-5-055005}.

The atom-molecule conversion process involves the study of second-order reactions for the case of molecular dimer \cite{yurovsky2006formation} :
\begin{center}
 $A + A \overset{\alpha}\leftrightharpoons A_{2}$.
\end{center}
The conventional chemical reaction process \cite{liu2017ultracold} takes place, when temperature is of the order of $100$K, with large number of particles at high-momenta. As mentioned earlier, the formation of cold molecules from atoms through PA/FR techniques has made it possible to induce chemical reaction in the condensed phase, with much fewer atoms at low-momenta. The mean field dynamics in the condensed phase is governed by a generalized non-linear Schr$\ddot{o}$dinger equation, wherein solitons and solitary waves are expected to be present in cigar shaped geometry. These excitations also manifest in several chemical reactions, the most well-known being the Belousov-Zhabotinsky reaction \cite{noyes1990mechanisms,gizynski2017chemical,hou2017bursting}, where atomic and molecular components exhibit spatio-temporal oscillations. Physically, conversion of atoms into molecules in `superchemistry' is the matter wave analogue \cite{al2011formation} of second harmonic generation of photons in non-linear crystals. Here, the possibility of coherent  production of   molecules with a high phase space density provides new possibilities in the field of molecular matter wave amplification \cite{meystre2004molecular,cheng2007quantum} and other non-linear phenomena.

A field theoretical approach for coupled atom-molecule BEC (AMBEC) has been developed, which includes pair correlations, quantum fluctuations and thermal effects \cite{drummond1998coherent,kheruntsyan1998multidimensional}. In this approach, dynamics of AMBEC is described by a modified coupled non-linear Schr\"odinger equation (NLSE). The difference from  pure two-species BEC \cite{ho1996binary,pu1998properties,park2000strong,haimberger2004formation} described by coupled NLSE, arises due to inter-conversion, originating from PA, which induces an additional quadratic non-linearity as compared to the cubic non-linearity in case of weakly coupled BEC \cite{Pethick2002}. Excitations like dark \cite{burger1999dark,denschlag2000}, gray \cite{khaykovich2002formation} or bright solitons \cite{alkhawaja2002bright,cornish2006formation}, vortices \cite{matthews1999vortices,dafovo1996bosons,abo2001observation} and Josephson-like oscillations \cite{smerzi1997quantum,albiez2005direct} are well studied in case of single as well as two-component BEC. Simultaneous appearance of cubic and quadratic interaction terms in AMBEC may provide novel cross-phase modulation \cite{PhysRevA.40.5063}, affecting the conversion process in the presence of PA.

Starting from an atomic condensate with controllably small fraction of condensed molecules, we show the complete conversion of condensed atoms into molecular BEC. This is achieved through a unique pair of solitonic excitations. These are unique in the sense that there are no periodic cnoidal counterparts. Interestingly, the molecular density, i.e., the grayness of molecular profile, governed by PA, which is also found to control the speed of the soliton. Gray soliton represents the moving reaction front, similar to the ones arising in chemical processes. The molecular dispersion profile shows Lieb mode like behaviour \cite{lieb1963} for small conversion rate. However, for large rate of conversion, degeneracy is found to set in. Surprisingly, when excitations are modulated with a plane wave phase, molecular dispersion exhibits bistablity, characterizing re-entrant behaviour. The plane wave momentum and PA govern the re-entrant dynamics, which can be controlled through phase manipulation.


\begin{figure}[t]
\begin{center}
\includegraphics[scale=0.4]{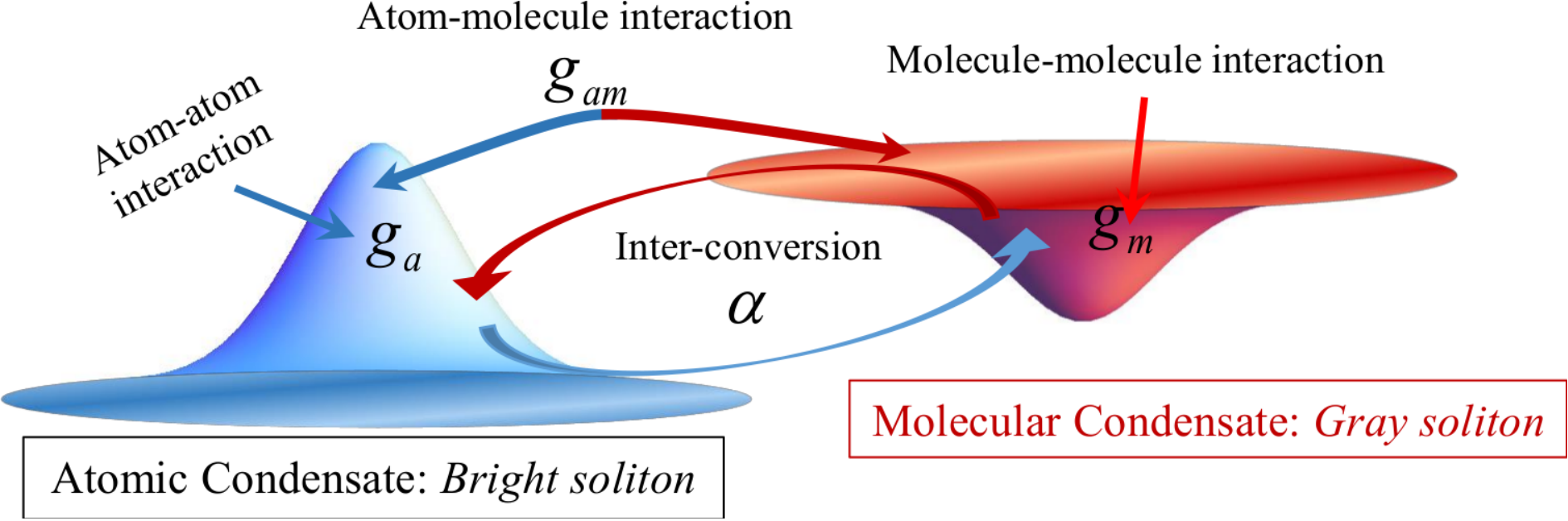}
\caption{ A schematic representation of the atomic Bose gas, coupled to a molecular condensate, described by bright and gray solitons.}
\label{fig1}
\end{center}
\end{figure}


The paper is organized as follows. In Sec.~\ref{sys}, we introduce the model and conserved quantities, followed by a detailed analysis of the unique bright-gray soliton pair in Sec.~\ref{solu}.  We investigate the dispersion of the molecular wave packet in Sec.~\ref{disp}, pointing out novel features not present in the single component gray soliton also known as the Lieb mode \cite{jackson2002}. The effects of conversion and momentum in bringing in a re-enovant phase is also demonstrated. We conclude in Sec.~\ref{conc} with directions for future work.

\section{System and Hamiltonian} \label{sys}

At low temperature, a weakly interacting BEC of atoms is described by the semi-classical Gross-Pitaevskii equation. These equations can be readily extended to include a molecular BEC, produced by coherent inter-conversion of atoms into molecules via PA;
\begin{eqnarray}
i \hbar \frac{\partial\phi_{a}}{\partial t} &=& -\frac{\hbar^2}{2m}\frac{\partial^{2} \phi_{a}}{\partial x^{2}} + (g_{a}|\phi_{a}|^{2} + g_{am} |\phi_{m}|^{2}) \phi_{a} \nonumber \\& & \hskip3.5cm + \sqrt{2} \alpha \phi_{m}\phi_{a}^{*}, \label{ac}  \\
i \hbar \frac{\partial\phi_{m}}{\partial t} &=& -\frac{\hbar^2}{4m}\frac{\partial^{2} \phi_{m}}{\partial x^{2}}+(\epsilon + g_{m}|\phi_{m}|^{2}+g_{am}|\phi_{a}|^{2})\phi_{m} \nonumber \\& & \hskip3.5cm + \frac{\alpha}{\sqrt{2}}\phi_{a}^{2} \label{mc}.
\end{eqnarray}
Here $\phi_{a}$ is the atomic BEC wavefunction and $\phi_{m}$ is that of molecules, with $g_{a}$, $g_{m}$ and $g_{am}$, measuring respective strengths of intra and inter-species interactions. This system is characterized by the conserved quantity,
\begin{eqnarray}
N = \int \big( |\phi_{a}|^{2} + 2 |\phi_{m}|^{2} \big) dx = N_{a} + 2 N_{m}.
\end{eqnarray}
Energy mismatch $\epsilon$ refers to the difference between the bound state energy of the two atoms to that of the energy of a free atom pair. We consider below a compatible soliton pair, which enables complete conversion of the atomic cloud to a molecular condensate.


\section{Solitonic excitations} \label{solu}


Solitons are known to play crucial role in case of excitation induced chemical reactions at low temperatures. Keeping in mind the possiblity of complete atom to molecule conversion, we consider the following bright-gray soliton pair,
\begin{eqnarray}
\phi_{a}(x,t) &=& \sqrt{\sigma_{0a}} \cos\theta ~\textrm{sech}\left[\frac{\cos \theta}{\zeta} (x - u t)\right] \label{ans1}
\label{ans2}
\end{eqnarray}
and
\begin{eqnarray}
\phi_{m}(x,t) &=& \sqrt{\sigma_{0m}} \sin \theta + i\sqrt{\sigma_{0m}} \cos\theta \nonumber \\& & \hskip 1cm \times \tanh\left[\frac{\cos\theta}{\zeta} (x - u t)\right],
\label{ans2}
\end{eqnarray}
here $\sigma_{0a}$ and $\sigma_{0m}$ are the amplitudes of bright and gray solitons of the condensate, with $u$ being the velocity. $\theta$ is the Mach angle and $\zeta$, the healing length of the solitonic profile. Direct substitution lead to the amplitudes of solitonic pair,


\begin{figure}[t]
\begin{center}
\includegraphics[scale=0.3]{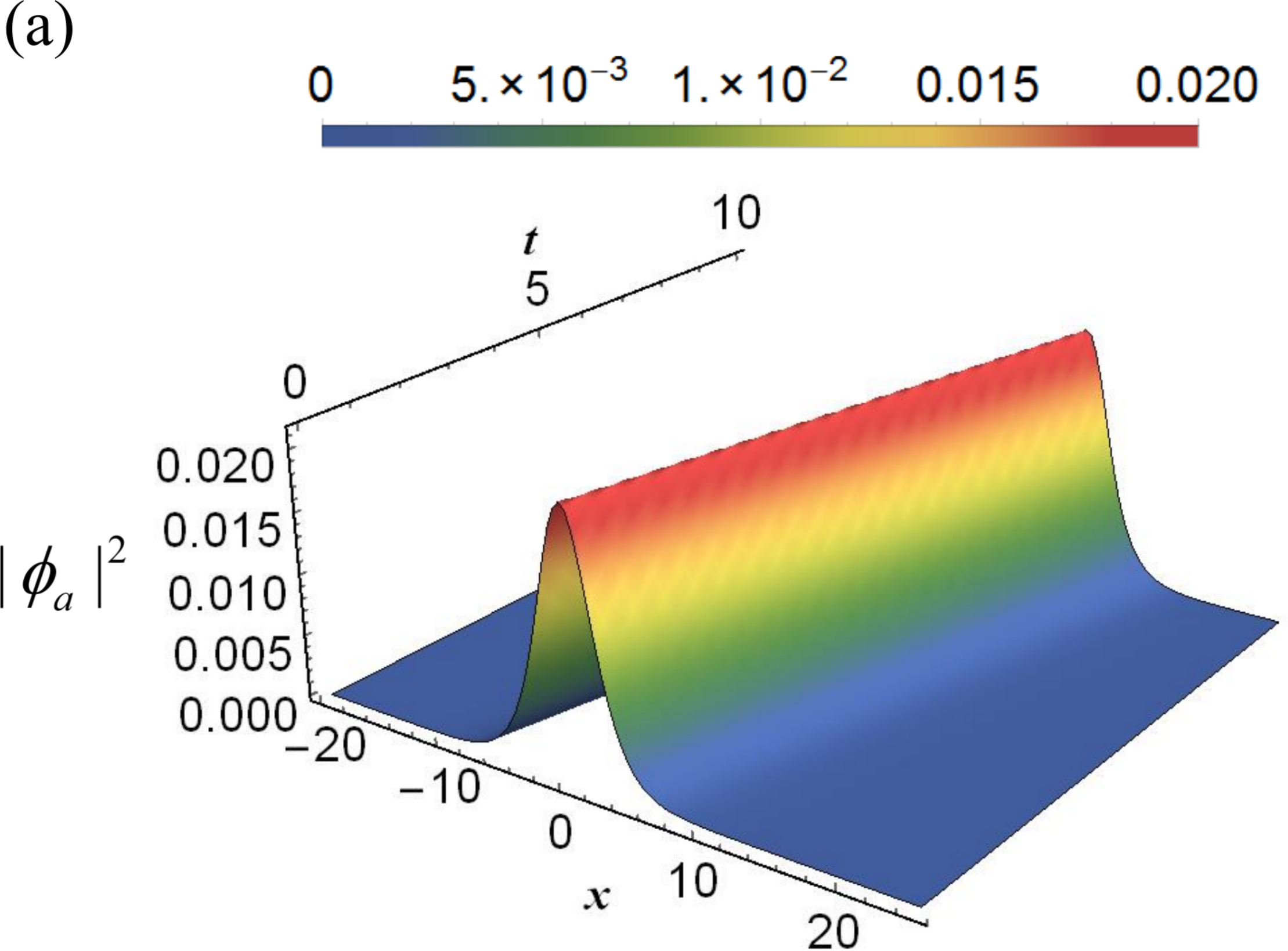}\\
\includegraphics[scale=0.3]{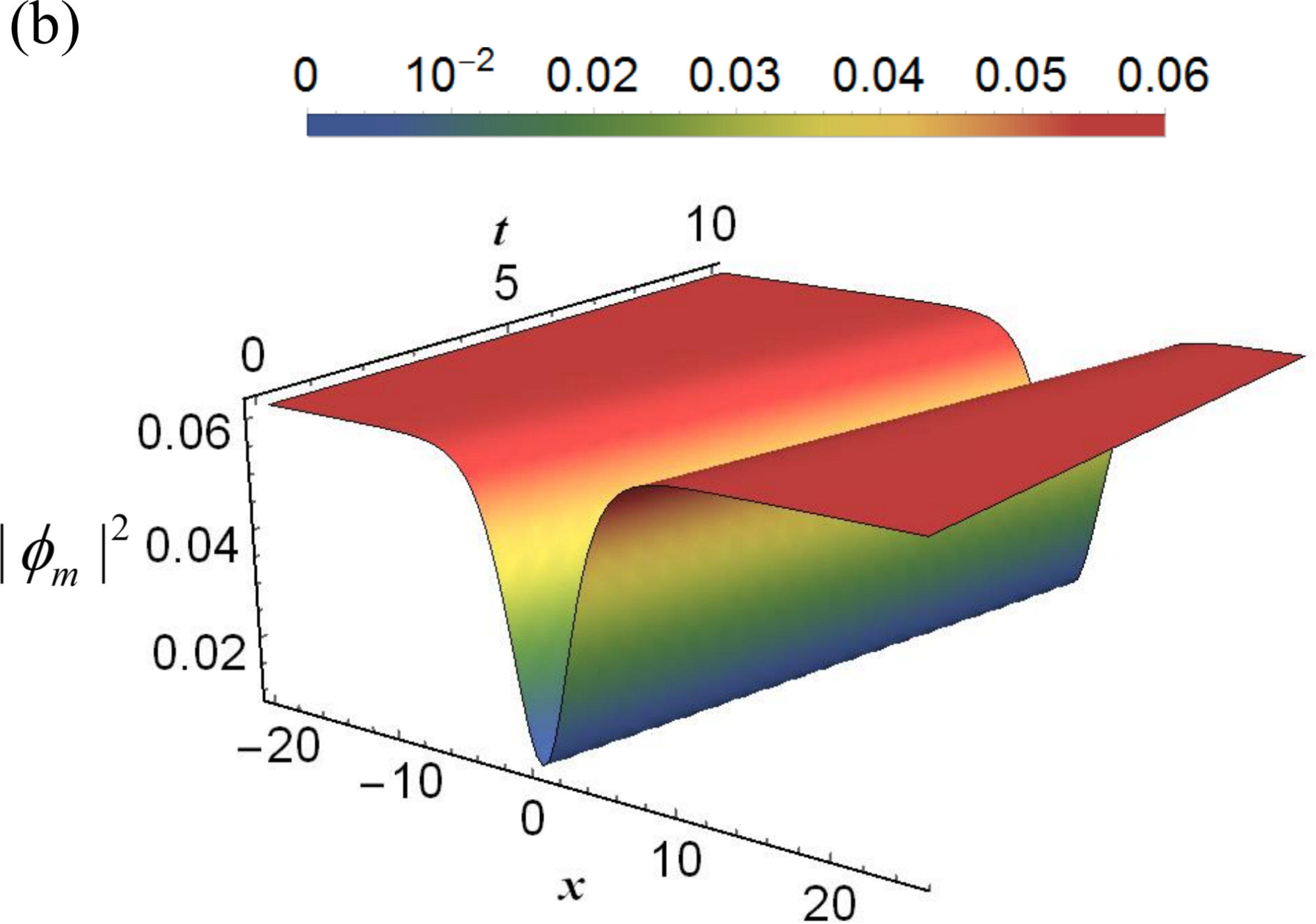}
\caption{ Depiction of the density profiles of atomic and molecular wave packets. (a) shows the propagation of bright soliton in the atomic condensate with that of complex gray soliton in the molecular condensate in (b). The parameter values are, $g_{a} = 3, g_{m} = 0.4, g_{am} = 1, \epsilon = -0.01, \alpha = 0.5$. }
\label{fig2}
\end{center}
\end{figure}

\begin{eqnarray}
\sigma_{0a} = \sigma_{0m} \frac{(g_{am}-2g_{m})}{(g_{a}-2g_{am})}, \qquad \sigma_{0m} = - \frac{\epsilon}{g_{m}},
\label{amp}
\end{eqnarray}
The energy mismatch is given by,
\begin{equation}\label{eps}
  \epsilon = - \frac{\hbar^{2} \cos^{2}\theta} {2 m \zeta} \frac{g_{m}}{g_{am}}.
\end{equation}
It is seen that the nature of molecule-molecule and atom-molecule interactions determine the sign of the energy mismatch. We consider the case where $\epsilon<0$ \cite{oles2007solitons} and obtain explicit solutions for general values of the couplings. Positivity of molecular density is ensured only when $g_{m}>0$. The atom-molecule interaction needs to be repulsive, $g_{am}>0,$ as is evident from Eq.~(\ref{eps}). A careful analysis reveals that the above possibility exists only within the following two parameter regimes: $g_{a}/2 > g_{am} > 2g_{m}$ or $g_{a}/2 < g_{am} < 2g_{m}$. The density profiles of both atomic (bright) and molecular (gray) solitons are shown in Fig.~(\ref{fig3}a) and (\ref{fig3}b), respectively. The healing length and the Mach angle are found to be
\begin{eqnarray}
\zeta &=& \frac{\hbar}{\sqrt{ 2m(g_{m} \sigma_{0m} - g_{am} \sigma_{0a} ) }}
\label{vel}
\end{eqnarray}
and
\begin{eqnarray}
 \sin \theta &=& \frac{u \gamma}{\sqrt{g_{\textrm{eff}}\sigma_{0m}/2m}},
\label{vel}
\end{eqnarray}
with $\gamma = \sigma_{0a}/2 \sigma_{0m} - 1$ and $g_{\textrm{eff}}=(g_{a}g_{m}-g_{am}^2) / (g_{a}-2g_{am})$.

It is evident that molecular density vanishes for $\theta = 0$, i.e., $\alpha=0$, implying only atoms are present in the absence of PA. During evolution, two atoms combine to form molecules, eventually leading to the formation of a molecular BEC. This limit is achieved at $\theta = \pi/2$, where the density of atomic condensate vanishes and molecular density attains its maximum, $|\phi_{m}|^{2} = \sigma_{0m}$. This unique pair of solitary excitations lead to complete conversion of atoms to molecular condensate. The parametric interaction and PA can be used to control how quickly the reactants are used up. This is in sharp contrast to the prediction of usual chemical kinematics, where rates do not depend on the number of participating particles and tend to zero at low temperature.

Healing length is governed by molecule-molecule and atom-molecule interactions unless $g_{m}<g_{am}\sigma_{0a}/{\sigma_{0m}}$. The velocity of the condensate is directly related to inter-conversion, $u = \alpha\sqrt{ 2 \sigma_{0m} }\zeta/\hbar$. Atom to molecule conversion takes its maximum when conversion reaches, $\alpha_{max} = g_{eff}\sqrt{\sigma_{0m}}/2\gamma$, with $u_{max} = \frac{1}{\gamma}\sqrt{g_{eff} \sigma_{0m}/2m}$. In the following, we study the dispersion of the molecular wave packet.

\begin{figure}[t]
\begin{center}
\includegraphics[scale=0.3]{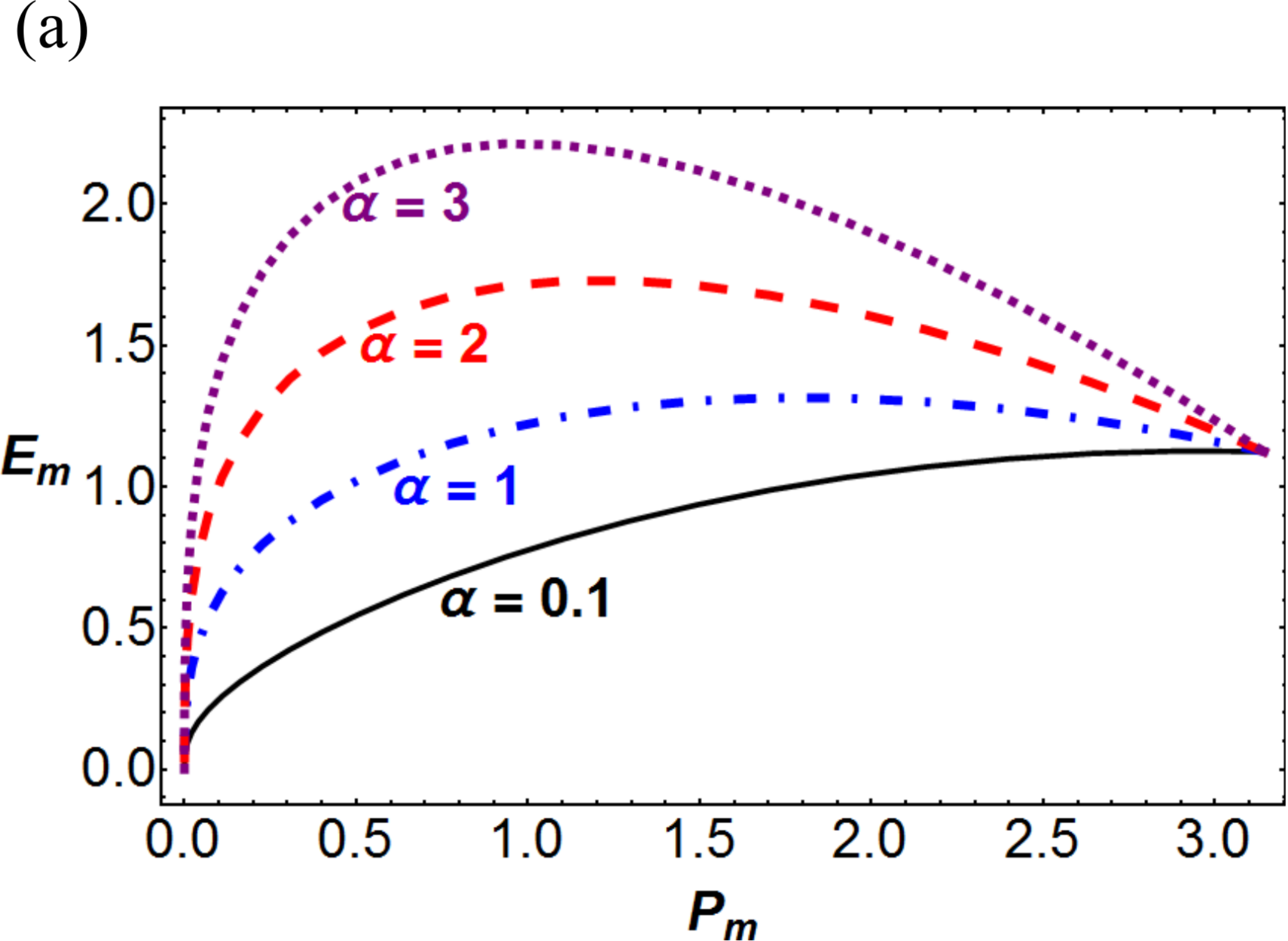}\\
\includegraphics[scale=0.3]{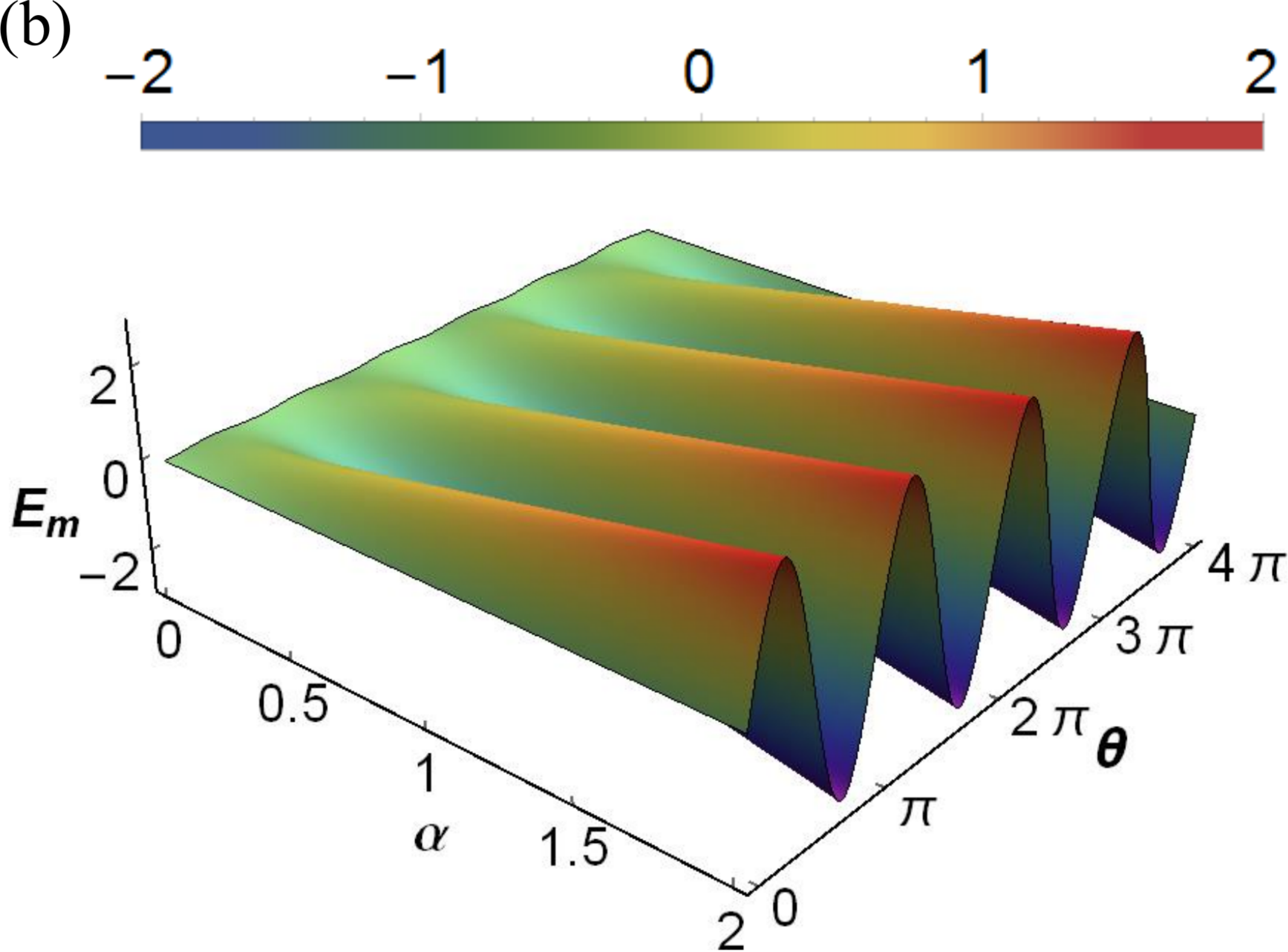}
\caption{(a) Shows dispersion relation of molecular condensate for different values of $\alpha$, revealing a degeneracy in molecular energy for higher values of $\alpha$. (b) depicts the variation of molecular energy as a function of the inter-conversion and Mach angle. The periodic nature of energy clearly shows the coherent nature of atom-molecule inter-conversion. Here parameter values are same as that in Fig. (\ref{fig2}). }
\label{fig3}
\end{center}
\end{figure}

\begin{figure}[t]
\begin{center}
\includegraphics[scale=0.28]{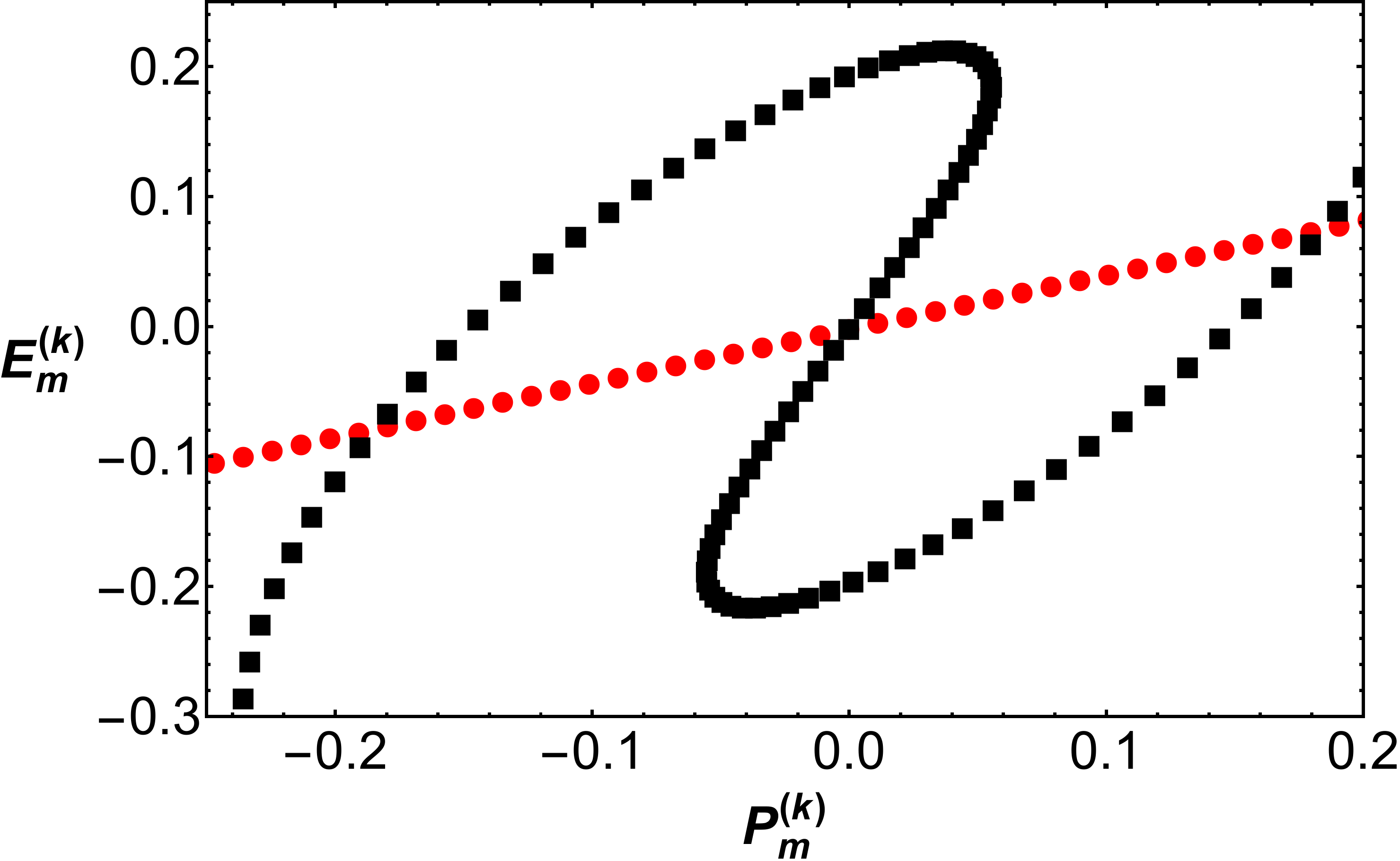}
\caption{The figure shows dispersion relation when the solitons are modulated by a plane wave phase. For small values of rate-conversion ($\alpha = 0.15$), a linear behavior is observed.  Higher values of $\alpha$ results in bistability, indicating phase re-entrance. The other parameter values are same as that in Fig. (\ref{fig2}).
}
\label{fig4}
\end{center}
\end{figure}


\section{Dispersion relation}\label{disp}

We consider the hydrodynamic variables defined by $\phi_{m} = \sqrt{\sigma_{m}(\mathcal{Z})} e^{i \chi_{m}(\mathcal{Z})}$ with $\mathcal{Z} = x - u t$, where density and phase of the molecular condensate satisfies,
\begin{eqnarray}
  \frac{\hbar}{m}\frac{\partial \chi_{m}(\mathcal{Z})}{\partial \mathcal{Z}} &=& u (1 - \frac{\sigma_{0m}}{\sigma_{m}})\\
  \sigma_{m}(\mathcal{Z}) &=& \sigma_{0m}  - \sigma_{0m} \cos^{2}\theta~ \textrm{sech}^{2}\left[\frac{\cos \theta}{\zeta} \mathcal{Z}\right].
\end{eqnarray}
For the soliton profiles in Eqs. (\ref{ans1}) and (\ref{ans2}), the energy of the molecular condensate yields,
\begin{eqnarray}
E_{m} &=& \frac{4}{3} g_{m} \sigma_{0m}^2 \zeta \cos^{3}\theta + 2 g_{am} \sigma_{0a} \sigma_{0m} \zeta (\cos\theta - \frac{2}{3}\cos^{3}\theta) \nonumber \\ & &\hskip1cm + \sqrt{2} \alpha \sigma_{0a} \sqrt{\sigma_{0m}} \zeta \sin2\theta.
\end{eqnarray}
The first and second terms arise due to the molecule-molecule and atom-molecule interactions and the last term originates from atom-molecule inter-conversion. The renormalized momentum of the complex gray soliton is given by,
\begin{eqnarray}
P_{m} &=& \hbar \sigma_{0m} \left(\pi \frac{u}{|u|} - 2 \theta - \sin2\theta \right).
\label{mom}
\end{eqnarray}
It is observed that the presence of inter-conversion affects both the energy and momentum of gray soliton as the Mach angle depends on $\alpha$. The energy-momentum dispersion is shown in Fig. (\ref{fig3}a). For small values of conversion, $\alpha = 0.1$, dispersion response coincides exactly with the Lieb mode in case of single component BEC \cite{jackson2002}. Interestingly, as one increases the rate of conversion, degeneracy sets in showing energy equality for two different values of momentum. With a higher value of $\alpha$, molecular energy increases as more atoms participate in forming molecules. Blue (dashed-dot), red (dashed) and purple (dotted) curves correspond to $\alpha = 1, 2 , 3$ respectively. Molecular momentum achieves its maximum, $P^{max}_{m} =\pi \hbar \sigma_{0m}$, at $\theta = 0$. Fig.~(\ref{fig3}b) shows the variation of molecular energy with the inter-conversion term and Mach angle. It is worth mentioning that the limiting case for $\alpha \to0$ and $\alpha=0$, show distinctly different behaviour. For $\alpha=0$, velocity of the gray soliton is found to be a free parameter \cite{jackson2002,das2016realization} whereas for $\alpha \to 0$, soliton velocity takes a vanishing value.

We now consider the case when condensate pair is locked with a plane wave phase, $\phi_{a} \rightarrow \phi_{a} e^{ikx}$ and $\phi_{m} \rightarrow \phi_{m} e^{2 ikx}$. Inclusion of such phase, turns the molecular momentum to $P_{m}^{(k)}=P_{m}-4\sigma_{0m}k \zeta cos\theta $, with a shift in energy, $E_{m}^{(k)}=E_{m}-8\sigma_{0m}k^2 cos\theta/\zeta$, exhibiting bistable behavior, shown in Fig. (\ref{fig4}). For small rate of conversion, $\alpha = 0.15$, dispersion shows linear profile, figured by red (circle) curve. In Fig. (\ref{fig4}) bistability appears for $\alpha = 2.25$. Such bistability has also been seen in case of polaritonic excitations in strongly nonlinear waveguide \cite{hafezi2012quantum,das2013realization}.

\section{Conclusion}\label{conc}
In conclusion, we have identified a unique pair of localized excitations for complete quantum state transfer, controlled by photoassociation in a coupled atom molecular BEC. The initial density of the molecular condensate is also controlled by the photoassociation and other couplings. Speed of the reaction front is also found to be regulated by photoassociation, results in complete conversion to a molecular condensate for a finite value of inter-conversion. The parameter $\alpha$ controlling photoassociation, leads to two different physical scenarioes in case of $\alpha=0$ and $\alpha\to 0$. In the absence of $\alpha$, soliton velocity remains a free parameter, whereas for low conversion rate, it takes a vanishing value. The molecular dispersion configuration is analogous to the Lieb mode, for small conversion and is significantly affected by photoassociation, where a degeneracy is observed for large values of $\alpha$. It is further observed that a plane wave modulation leads to a re-entrant phase. Solitons have found application in information storage and retrival \cite{agarwal2003slow,jing2011quantum}. We hope that this control pair of soliton may find similar applications in chemical quantum memory \cite{jing2011quantum,reim2010towards,allen2012optical}.

\section*{Acknowledgement}
PD acknowledges the financial support provided by the Board of Research in Nuclear Sciences, Department of Atomic Energy (DAE), Government of India, under the project no. RP03178.

\section*{References}

\bibliography{ms}

\end{document}